\documentclass[review]{elsarticle}

\usepackage{lineno,hyperref}
\usepackage{amsmath}
\usepackage{mathtools}
\usepackage{amssymb}
\usepackage{subfig}
\modulolinenumbers[1]

\graphicspath{{figures/}}
\newcommand{\dt}{\Delta t}
\newcommand{\bl}{\begin{linenomath}}
\newcommand{\el}{\end{linenomath}}
\newcommand{\be}{\begin{equation}}
\newcommand{\ee}{\end{equation}}

\DeclareMathOperator\diag{diag}
\DeclareMathOperator\RMSE{RMSE}
\renewcommand{\vec}[1]{\boldsymbol{#1}}
\renewcommand{\tilde}{\widetilde}

\usepackage{xcolor}
\definecolor{backgreen}{rgb}{0.00, 0.169, 0.212}
\definecolor{textgray}{rgb}{0.514, 0.580, 0.589}
\newcommand{\aside}[1]{\textcolor{black}{#1}}
\newcommand{\defeq}{\coloneqq}

\journal{Advances in Water Resources}

\begin{document}


\begin{frontmatter}

\title{On the accuracy of simulating mixing by random-walk particle-based mass-transfer algorithms \tnoteref{mytitlenote}}
\tnotetext[mytitlenote]{The authors were supported by the National Science Foundation under awards EAR-1417145 and DMS-1614586.}


\author{Michael J. Schmidt, Stephen D. Pankavich}
\address{Applied Mathematics and Statistics, Colorado School of Mines, Golden, CO 80401, USA }

\author{David A. Benson\corref{mycorrespondingauthor}}
\address{Hydrologic Science and Engineering, Colorado School of Mines, Golden, CO 80401, USA}

\cortext[mycorrespondingauthor]{Corresponding author}

\begin{abstract}
Several algorithms have been used for mass transfer between particles undergoing advective and macro-dispersive random walks. The mass transfer between particles is required for general reactions on, and among, particles. The mass transfer is shown to be diffusive, and may be simulated using implicit, explicit, or mixed methods. All algorithms investigated are accurate to $\mathcal{O} (\dt)$. For $N$ particles, the implicit and semi-implicit methods require inverse matrix solutions and $\mathcal{O} (N^3)$ calculations. The explicit methods use forward matrix solves and require only $\mathcal{O} (N^2)$ calculations. Practically, this means that naive implementations with more than about 5,000 particles run more reliably using explicit methods.

\end{abstract}


\end{frontmatter}


\section{Introduction}
The random-walk particle-tracking (RWPT) method was originally developed to simulate advective and dispersive transport of conservative or simply (linearly, instantaneously reversible) sorbing solutes \cite{labolle,Salamon_2006}. The method is attractive because it does not suffer from numerical dispersion or negative concentrations.  The method was extended \cite{Benson_react} to nonlinearly interacting (bimolecular) chemical reactions by sequentially calculating the product of the probabilities of particle collision and thermodynamic reaction. The actual reactions were then performed using Monte Carlo methods and particles were ``born'' or ``killed'' by a comparison of reaction probability to randomly-generated numbers.  The method was originally restricted to one, or a series of, bimolecular reactions \cite{Paster_WRR,Paster_JCP,dong_awr,Ding_monod,Ding_WRR,Benson_AWR_2016,Bolster_mass}, because any particle was composed of only one chemical species.  If the reaction is viewed as a mixing process, which may be denoted $2pA+2qA\rightarrow(p+q)A+(p+q)A$, then particles can carry as many species as desired, and mass transfer of all species occurs between particles \cite{Benson_arbitrary}.  The mass transfer still only occurs between particles with some probability of collision, and these probabilities may be viewed as the weights associated with mass transfer.  {\em Benson et al.} \cite{Benson_arbitrary} suggested that this collision-weighted mass transfer process follows a diffusion equation, although this was not shown rigorously.  Furthermore, those authors chose a particular explicit mass transfer scheme, while later studies used an implicit scheme \cite{Engdahl_WRR}.  Because both explicit and implicit schemes appear to work, it is plausible that a combination of these, similar to the Crank-Nicolson (C-N) algorithm, may increase accuracy.  \aside{The purpose of this paper is to first develop a framework to investigate whether the ``action'' of the mass-transfer algorithm proposed by {\em Benson et al.}  \cite{Benson_arbitrary} is actually diffusive.  Once this diffusive nature is shown,  the convergence rates of the several algorithms that immediately present themselves can be demonstrated. }

\section{Semi-implicit scheme}\label{sec_SI}

Among a total of $N$ particles located at positions $x_i$, the collision-weighted mass exchange over a time step $\dt$ is written
\aside{\bl
\be
m_j^{k+1} - m_j^k = \sum_{i=1}^N \frac12 \left(m_i^{k + \ell}-m_j^{k + \ell} \right) P \left(\left\vert x_i-x_j\right\vert; \dt\right),
\label{eq_transfer}
\ee
\el}
\noindent where the superscript denotes timestep (i.e., $m_j^{k} = m_j(k\dt))$, \aside{$\ell = 0, 1$}, and $P_{ij} = P \left(\left\vert x_i-x_j\right\vert; \dt\right)$ is the probability of particle collision.
\aside{This collision probability is shown to depend only upon the distance between particles, though it may have a more complicated form if non-isotropic or position dependent diffusion/dispersion paradigms are considered. Nonetheless, while the functional form of $P$ may change, the mass transfer algorithm would be unaltered.}
For particles undergoing Brownian motion, this is the convolution of each particle's Gaussian location density, which is also Gaussian (see \cite{Benson_react,Paster_JCP}).  If \aside{$\ell = 1$}, the calculation is implicit, and if \aside{$\ell = 0$}, the calculation is explicit (which may take several forms, for example, sequentially calculated or simultaneously calculated). A semi-implicit form is reminiscent of the Crank-Nicolson scheme and uses equal amounts of $k$ and $k+1$ masses, so that we may write \eqref{eq_transfer} as
\bl
\be\label{eq_CN}
m_j^{k+1} - m_j^k = \alpha\sum_{i=1}^N\frac12 \left(m_i^{k+1}- m_j^{k+1} \right) P_{ij} +(1-\alpha)\sum_{i=1}^N \frac12 \left(m_i^k-m_j^k\right)P_{ij},
\ee
\el
\noindent which uses $\alpha$=1, 1/2, and 0 for implicit, semi-implicit, and explicit formulations respectively. Now denote the masses as a vector, i.e., $\vec m = [m_1, \dots, m_N]^T$, and if one constructs a matrix of particle collision probabilities $\vec P$ with entries $P_{ij}$, then \eqref{eq_CN} can be expressed as
\bl
\be\label{eq_matrix}
 \left[ \vec I +\frac{\alpha}{2} \left(\diag\left( \vec 1 \vec P\right) -  \vec P\right) \right] \vec m^{k+1} = \left[ \vec I - \frac{1-\alpha}{2} \left(\diag\left( \vec 1 \vec P \right) + \vec P\right) \right] \vec m^k
\ee
\el
\noindent where \aside{$\vec A = \diag(\vec x)$} denotes a diagonal matrix, \aside{$\vec A$} with the entries of \aside{vector} $\vec x$ along the main diagonal and $\vec 1$ is an $1 \times N$ vector of ones.

\section{Explicit schemes}\label{sec_EX}
Clearly, setting $\alpha=0$ in \eqref{eq_matrix} results in an explicit forward matrix calculation. We call this matrix-explicit.  All of the masses used to calculate the transfer magnitudes are from the beginning of the timestep.  Another method sequentially calculates \eqref{eq_CN} for $j=1,\dots, N$. After the $j^{\text{th}}$ particle is updated, its new mass can be used on the right side of the equation for subsequent calculations. If the sum is calculated using one \aside{fixed} value for $m_j$, then we call this vector-explicit, \aside{calculated as followed (employing pseudo-code, where $\circ$ denotes the entry-wise, or Hadamard, product)}
\bl
\aside{\begin{equation}
    \begin{aligned}
        &\textbf{\texttt{for}}\ j = 1 : N\\
            & \qquad \vec{\Delta m} = \frac{1}{2} \left(\vec m(t) - m_j(t)\right) \circ \vec P_{(:, j)} \\
            & \qquad \vec m(t + \dt) =  \vec m(t) - \vec{\Delta m} \\
            & \qquad m_j(t + \dt) =  m_j(t) + \sum \vec{\Delta m} \\
        &\textbf{\texttt{end}}.
    \end{aligned}
\end{equation}}
\el
Furthermore, if the sum is expanded, then each calculation may use an updated $m_j$ accounting for all previous terms in the sum. We call this explicit-sequential, and it is \aside{calculated as follows}
\bl
\aside{\begin{equation}
    \begin{aligned}
        &\textbf{\texttt{for}}\ i = 1 : N\\
            &\qquad \textbf{\texttt{for}}\ j = 1 : N\\
                & \qquad \qquad \Delta m = \frac{1}{2} \left(m_i(t) - m_j(t)\right) \vec P_{(i, j)} \\
                & \qquad \qquad m_i(t + \dt) =  m_i(t) - \Delta m \\
                & \qquad \qquad m_j(t + \dt) =  m_j(t) + \Delta m \\
            &\qquad \textbf{\texttt{end}}\\
        &\textbf{\texttt{end}}.
    \end{aligned}
\end{equation}}
\el
This method has a computational advantage in that there is no matrix multiplication required (just two loops over particle numbers), and hence it can accommodate huge particle numbers.  It turns out that the vector-explicit algorithm is unstable for all ranges of parameters tested here and will not be explored further.  

\section {Accuracy as a function of repeated operation}

In general, the particle positions change due to non-uniform and potentially unsteady mean velocity.  The particles are also typically given a random component to represent diffusion and hydrodynamic dispersion; therefore, each simulation in an ensemble has subtle differences \cite{labolle}.  This is one advantage of the method: the evolving particle \aside{spacings (controlled by the number of particles) and masses} represent the heterogeneity of concentrations\aside{---as defined by evolving auto- and cross-correlation functions---}and the resulting mixing process \cite{Paster_JCP,Bolster_mass,Benson_AWR_2016}.  However, in order to check accuracy and convergence in this paper, we must artificially remove the randomness of simulations.  This is done by eliminating the random movements of particles and spacing them evenly on the interval $(0,1)$, where the number of particles dictates the size of the constant spacing.  This also allows us to construct the \aside{classical Eulerian} implicit finite-difference (FD) approximation of diffusion using a 3-point space stencil for comparison.  (We stress that our particle collision method may not be the most efficient way to simulate diffusion on a fixed grid of points, but the method will continue to work no matter how ``mixed-up'' the particle positions become.)

We track errors over time as functions of $N$, $\dt$ \aside{[T]}, and total time $k\dt$.  In all simulations we choose a diffusion coefficient $D=10^{-3}$ \aside{[L$^2$ T$^{-1}$]} and a total simulation time of $10$ seconds (unless specified otherwise).  For an initial condition (IC) we choose a Heaviside function to represent the most unmixed (and error-inducing) possible state.  We also choose a Gaussian IC to determine if errors remain more stable over time.  Our measure of error between simulations and analytic solutions uses the root-mean-square error (RMSE),
\bl
\begin{equation*}
    \RMSE(\vec s - \vec a) = \left(\frac1N \sum_{j=1}^N (s_j-a_j)^2\right)^{1/2},
\end{equation*}
\el
where $s_j$ and $a_j$ denote simulated and analytic solutions at spatial point $j$.  We also utilized the infinity norm, $ \max_j\left(\left\vert s_j-a_j\right\vert\right)$, which showed similar scaling and is not shown here for brevity.

\begin{figure}[t]%
    \centering
    \subfloat[]{\includegraphics[width=0.5\textwidth]{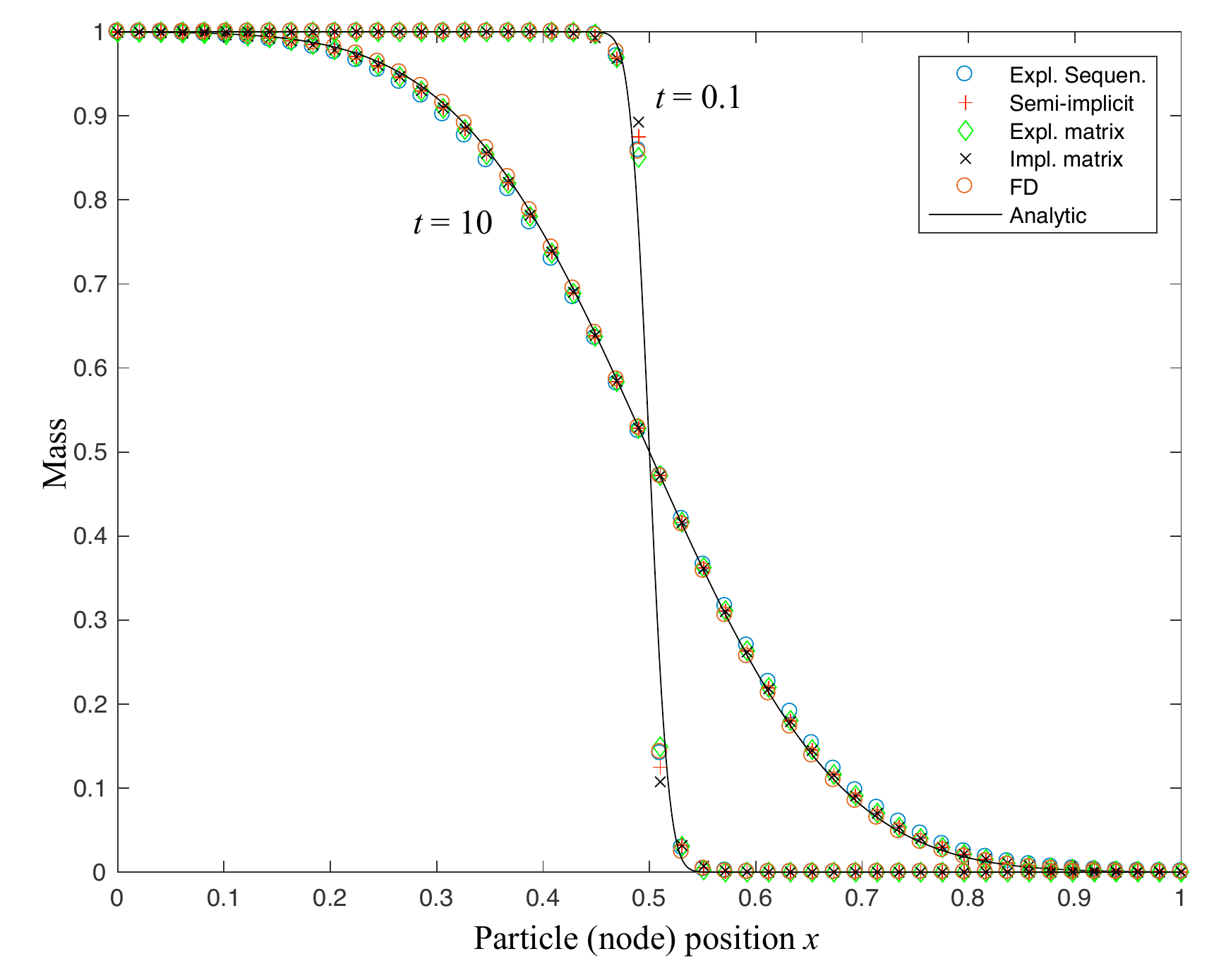} }%
    \subfloat[]{\includegraphics[width=0.5\textwidth]{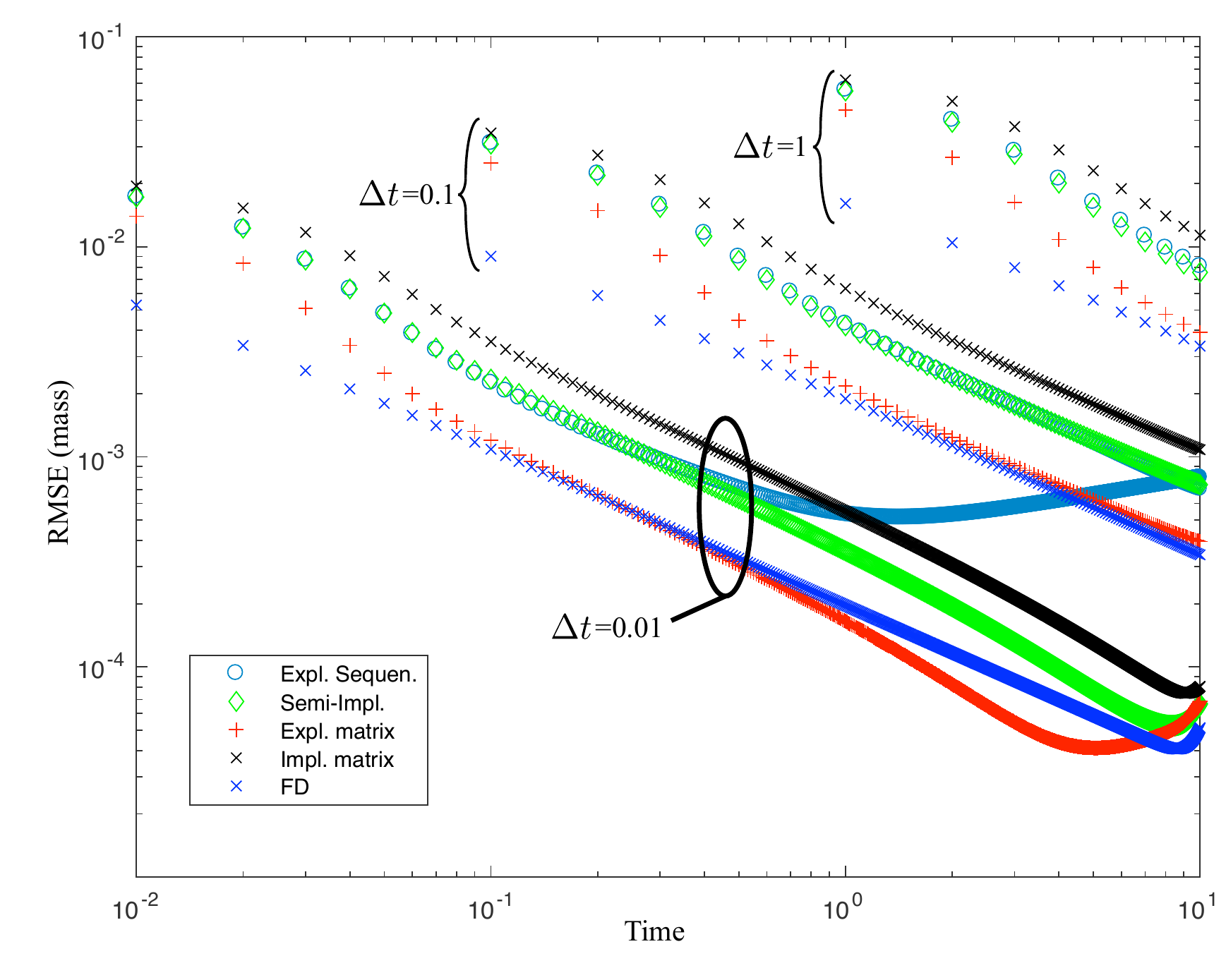} }%
    \caption{(a) Numerical approximations via particle mass-transfer and finite-difference (symbols) versus analytic solution (solid curve) at $t=0.1$ and $t=10$ with $\dt=0.1$, $D=10^{-3}$ and Heaviside IC. (b) RMSE from various methods over time for different values of $\dt$  with Heaviside IC.}%
    \label{fig:erf_vary_dt}%
\end{figure}

\begin{figure}[t]
\centering
\includegraphics[width=\textwidth]{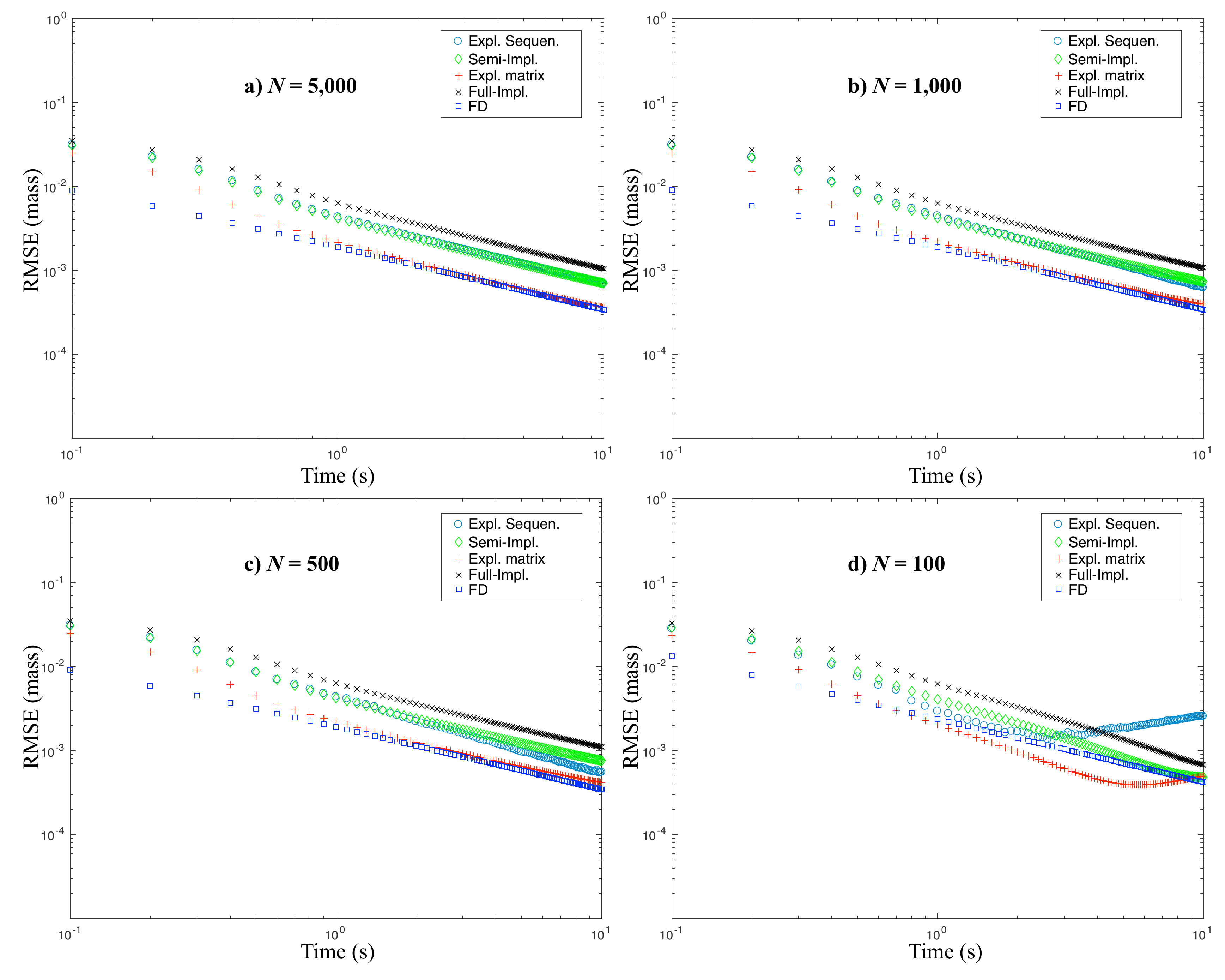}
\caption{RMSE from various methods over time for different number of particles (or spatial discretization) with Heaviside IC.}
\label{fig_vary_N}
\end{figure}

To illustrate the motivation for this technical note, for $N=50$ we see that all solutions appear diffusive by visual inspection of the plots of $m(x,t=10)$ (Fig. \ref{fig:erf_vary_dt} (a)). On the other hand, considering the various solution methods after one time step (here $\dt=0.1$), it is clear that the methods differ significantly in their ``one-step'' approximation of diffusion.  To isolate error incurred by time discretization, we first fix $\dt=0.1$ and vary the number of particles (Fig. \ref{fig_vary_N}).
\aside{The errors are similar for $N=500, 1000$, and $5000$, indicating that, as long as a sufficient, minimum number of particles is used, increasing particle number does not appreciably decrease error.}
In subsequent simulations we use $N=1000$ for consistency.  All methods achieve their greatest error at the beginning of the simulation, due to the unmixed, or infinite gradient, IC.  Repeated applications of the operators result in reduced error.  In other words, repeated application of the matrix operations converges to a true diffusive operator.  This is discussed further in Section \ref{sec_diff}.
\aside{Also evident on the plot is the relatively poor performance of both implicit and semi-implicit particle methods, relative to the explicit matrix particle method that tends to converge quickly to the accuracy of the Eulerian finite-difference solution to which we compare. } 

\section{Accuracy as a function of $\Delta t$}\label{sec_dt_acc}

For a given number of particles (here $N=1000$), the overall errors of all methods decrease over repeated application.
\aside{However, we note that the mass-transfer algorithm, investigated here, is only one component of a particle-tracking simulation that may involve other processes like diffusive random walks, advective motion, and chemical reaction. If these other processes are included, it may negate this property.}
One might expect that, similar to the Crank-Nicolson time-stencil in an FD implementation, the semi-implicit solutions would improve as $\dt$ decreases, relative to the explicit and implicit methods, but this is not the case.  All methods tested here have errors approximately proportional to $\dt$ (Fig. \ref{fig:erf_vary_dt} (b)).

\begin{figure}[t]%
    \centering
    \subfloat[]{\includegraphics[width=0.5\textwidth]{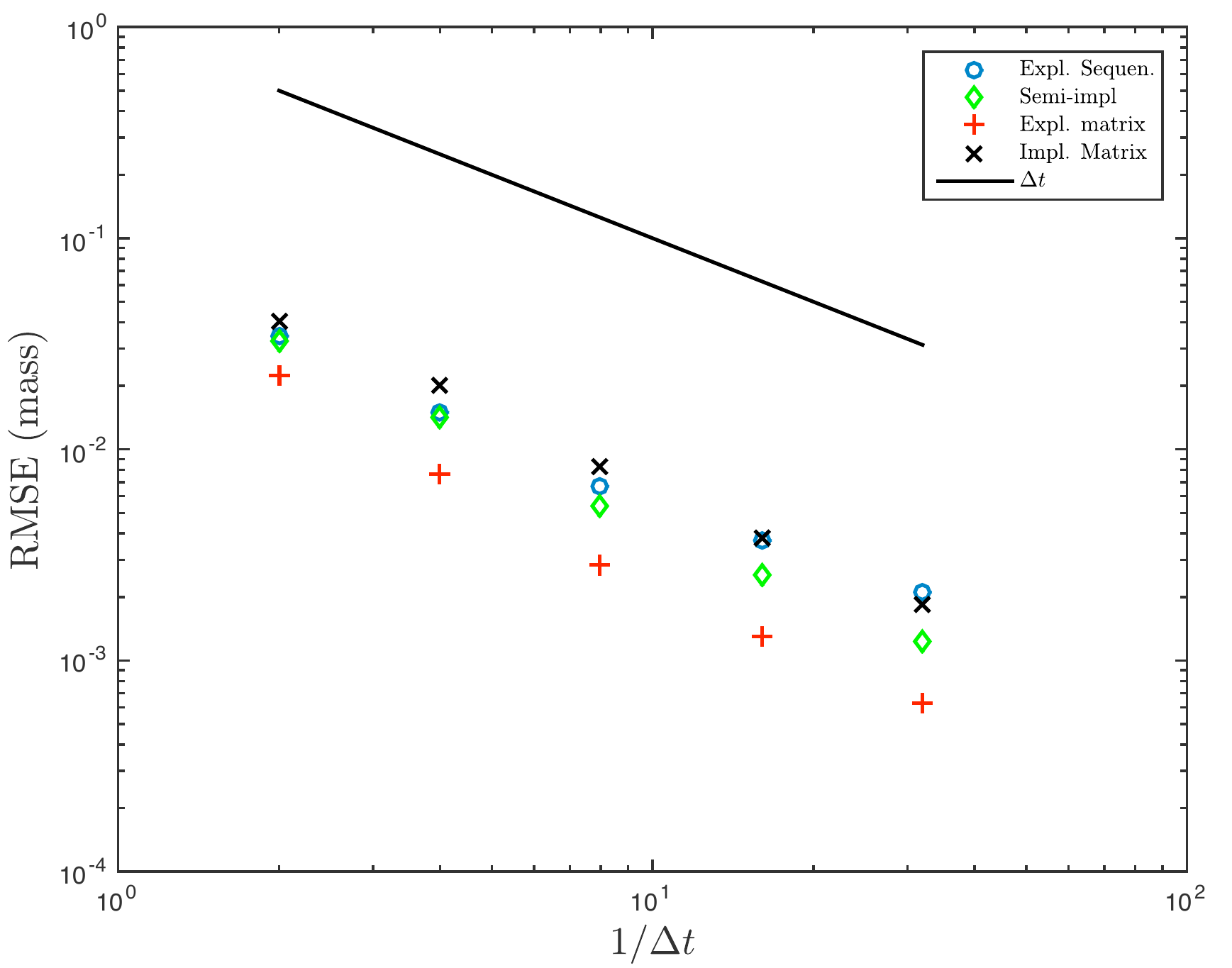} }%
    \subfloat[]{\includegraphics[width=0.5\textwidth]{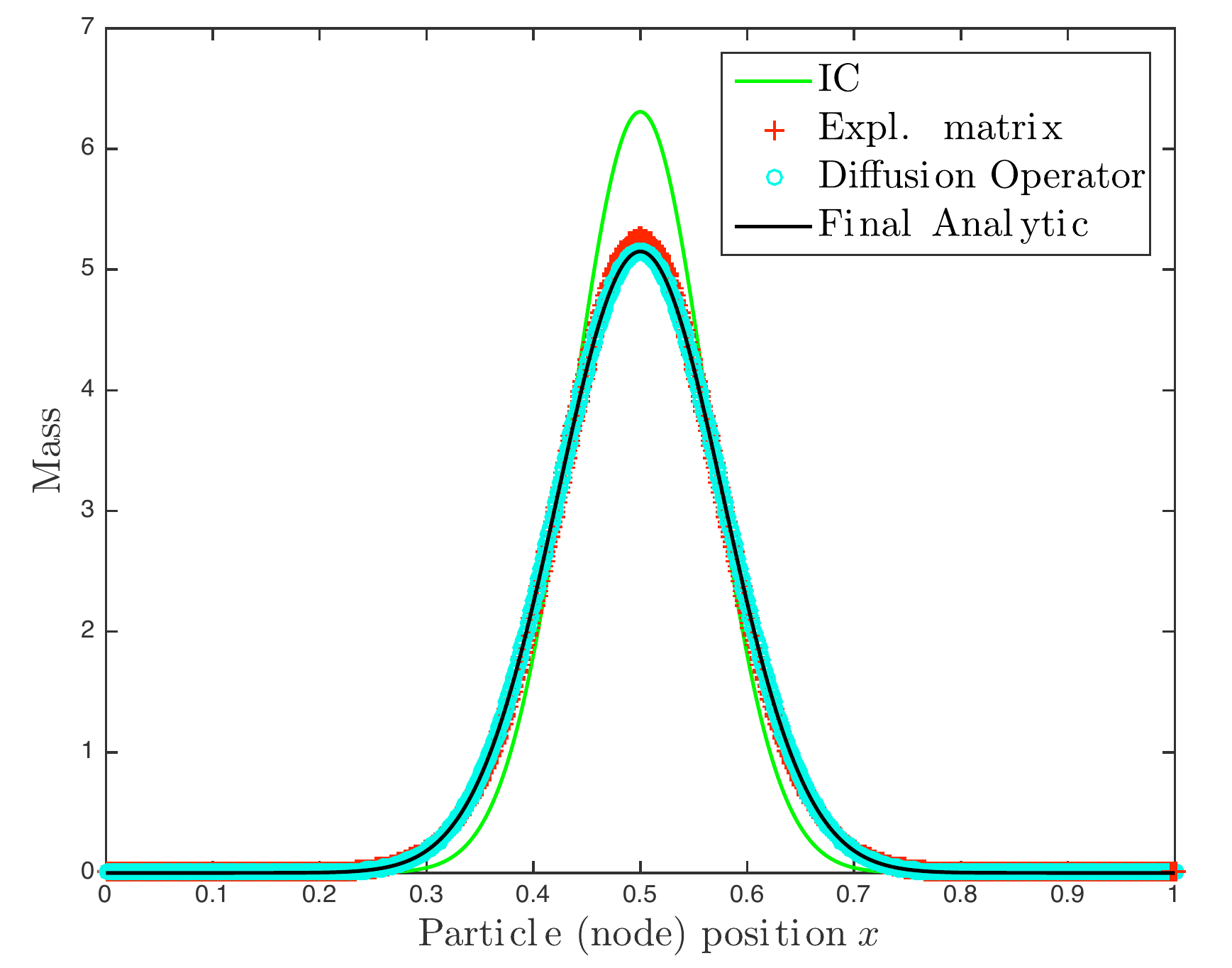} }%
    \caption{(a) RMSE vs. $1 / \dt$ showing first-order error decay in $\dt$ with Heaviside IC. (b) Plot of Gaussian IC and final simulated solutions \aside{$\left(\dt = 0.5\right)$} for explicit matrix method and discretized diffusion operator.}%
    \label{fig:EOC_gaussian}%
\end{figure}

To better understand the relation between error and $\dt$, we wish to find the power $p$ such that $\mathcal{E}_{A}\defeq \RMSE(\vec s - \vec a) < c(\dt)^{p} = \mathcal{O}(\dt^p)$, given the simulated and analytic solution vectors ($\vec s$ and $\vec a$) and some constant $c$.
Conducting a convergence analysis for a one-second simulation and refining $\dt$ by successive halves, we compute an experimental value of $p$, the estimated order of convergence (EOC) such that
\bl
\begin{equation*}
    \text{EOC} \defeq \frac{\frac{\log\mathcal{E}^{\text{old}}}{\log\mathcal{E}^{\text{new}}}}{\frac{\dt^{\text{old}}}{\dt^{\text{new}}}} = \left(\frac{1}{2}\right)\frac{\log\mathcal{E}^{\text{old}}}{\log\mathcal{E}^{\text{new}}}.
\end{equation*}
\el
For the Heaviside IC case, we see, in Table \ref{tab:convergence_analysis}, demonstrated first-order convergence in $\dt$ for all the discussed matrix methods.
A plot of these errors is shown in Fig. \ref{fig:EOC_gaussian} (a) with a reference line showing $\dt$.
As well, we see in Table \ref{tab:convergence_analysis} that the explicit sequential method suffers in accuracy for small $\dt$ and does not attain asymptotic convergence of $\mathcal{O}(\dt)$; this is visually depicted in Fig. \ref{fig:erf_vary_dt} (b) for $\dt = 0.01$.


\begin{table}[t]
    \caption{Convergence analysis of mass-transfer algorithms to analytic solution, Heaviside IC.}
    \label{tab:convergence_analysis}
    \centering

    \begin{tabular}{|l|ll|ll|ll|ll|}
    \hline

    \hline
    & \multicolumn{2}{c|}{\textbf{Expl. Seq.}} & \multicolumn{2}{c|}{\textbf{Semi-Impl.}} & \multicolumn{2}{c|}{\textbf{Expl. Mat.}} & \multicolumn{2}{c|}{\textbf{Full-Impl.}} \\
    \hline
    \textbf{$\dt$} & \textbf{$\mathcal{E}_{A}$} & \textbf{EOC} & \textbf{$\mathcal{E}_{A}$} & \textbf{EOC}& \textbf{$\mathcal{E}_{A}$} & \textbf{EOC}& \textbf{$\mathcal{E}_{A}$} & \textbf{EOC} \\
    \hline
         1    & 0.0338 &        & 0.0327 &        & 0.0222 &        & 0.0408 &  \\
         1/2  & 0.0146 & 1.2058 & 0.0141 & 1.2122 & 0.0076 & 1.5483 & 0.0203 & 1.0059 \\
         1/4  & 0.0067 & 1.1265 & 0.0054 & 1.3719 & 0.0028 & 1.4365 & 0.0082 & 1.2975 \\
         1/8  & 0.0036 & 0.8821 & 0.0025 & 1.1030 & 0.0013 & 1.1064 & 0.0037 & 1.1259 \\
         1/16 & 0.0022 & 0.6939 & 0.0012 & 1.0339 & 0.0006 & 1.0463 & 0.0018 & 1.0333 \\
    \hline

    \hline
    \end{tabular}
\end{table}

\section{Convergence to a diffusive process}\label{sec_diff}

In an infinite 1-D domain, \aside{the solution to the diffusion equation at time $T$ is}
\bl
\aside{\begin{equation}
    \begin{aligned}
        m(x, T) &= (G \star m) (x, T)\\
        &= \int_{\mathbb{R}} G(x - x_0, T) m(x_0, 0) dx_0,
    \end{aligned}
\end{equation}}
\el
where \aside{$G(x, t) = \left(4 \pi D t\right)^{-1/2}\exp\left[-x^2/(4 D t)\right]$} is the Green's function for the diffusion equation, and $\star$ denotes convolution.
For time-discretized simulations, this convolution operation may be applied to the initial condition \aside{$k = T / \dt$} times using \aside{$G(x, \dt)$} to generate \aside{$m(x, T)$}.
In a space-discretized particle case where our initial condition is composed of $N$ Dirac deltas, each with position $x_i$ and mass $m_i$, this may be denoted \aside{$\vec m(t = T) = \left[{\vec D}\right]^k \vec m(0)$}, where \aside{$\left[\vec D\right]^n$ is an $n$-fold matrix product},
\aside{$\tilde{\vec D}_{ij} = G(\left\vert x_i - x_j \right\vert, \dt)$} and $\vec D$, our diffusion operator, is the result of normalizing the columns of $\tilde{\vec D}$, in order to preserve mass.
For suitable IC, the diffusion operator is virtually error-free, as compared to the analytic solution, and error can be driven to machine precision with a sufficient level of discretization (sufficiently large $N$, in the particle case).
However, the Green's function used to generate $\vec D$ assumes an infinite domain, and as a result is highly sensitive to boundary effects, as compared to the various mass-transfer algorithms developed in Section \ref{sec_SI} that naturally handle any boundary conditions since there are no particles to interact with outside the boundary.

While none of the typical numerical approximations (e.g., finite-difference, finite-element) are exactly diffusive, in that their matrix operator is exactly equivalent to $\vec D$, it suffices to show that, after $k$ applications of their matrix operator to $\vec m(0)$, the differences in \aside{$\vec m(T)$} are small.
In other words, if some process \aside{$\vec m(T) = {\vec A}^k \vec m(0)$} is ``diffusive'', then repeated applications have $\mathcal{E}_D \defeq \RMSE \left({\vec A}^k \vec m(0) - {\vec D}^k \vec m(0)\right) < \epsilon$ for some level of error, $\epsilon > 0$.

For this analysis, we will only consider the explicit matrix algorithm (i.e., Eq. \eqref{eq_matrix} with $\alpha=0$), as it consistently shows the lowest error of all described methods, and all matrix algorithms are consistently $\mathcal{O}(\dt)$.
Additionally, in order to avoid undesirable boundary effects experienced by the diffusion operator, a domain-centered Gaussian IC was used in favor of the Heaviside IC employed previously.
A plot of the initial condition and final solutions for the explicit matrix and diffusion operator algorithms is shown in Fig. \ref{fig:EOC_gaussian} (b) for $\dt = 0.5$ and one second of simulation time.
Again, performing a convergence analysis as in Section \ref{sec_dt_acc}, we see first-order convergence in $\dt$ of our algorithm to the discretized diffusion operator, as shown in Table \ref{tab:convergence_analysis_diff}.


\begin{table}[t]
    \caption{ Convergence analysis of explicit matrix algorithm to diffusion operator, Gaussian IC.}
    \label{tab:convergence_analysis_diff}
    \centering

    \begin{tabular}{|l|ll|}
    \hline

    \hline
    \textbf{$\dt$} & \textbf{$\mathcal{E}_D$} & \textbf{EOC} \\
    \hline
         1    & 0.0362 & \\
         1/2  & 0.0178 & 1.0241 \\
         1/4  & 0.0088 & 1.0097 \\
         1/8  & 0.0044 & 0.9941 \\
         1/16 & 0.0022 & 1.0064 \\
    \hline

    \hline
    \end{tabular}
\end{table}

\section{Discussion and Summary}
In this technical note we show that the inter-particle mass transfer algorithm can be simulated in implicit, semi-implicit (pseudo-Crank-Nicolson), and several explicit methods.  All have error that scales with $\mathcal O (\dt)$.  The matrix-explicit algorithm has the best performance in terms of both error magnitude and computational requirement, \aside{only requiring a matrix-vector multiplication (as opposed to matrix inversion for the implicit and semi-implicit methods)} of $\mathcal O (N^2)$ operations. \aside{Additionally, this computational cost can be lowered further if sparse linear algebra methods are employed, because, in practice, co-location probability is often considered to be zero for particles separated by distances greater than a few standard deviations of the diffusion process.}  We also show that, for an infinite domain, a simple convolution with the diffusion kernel has low error and effort of $\mathcal O (N^2)$.  However, this method suffers error if natural boundaries exist, because the kernel changes shape especially near the boundaries. \aside{For the sake of brevity, only 1D results have been presented in this text. However, 2D results for an analogous mass transfer algorithm are discussed in \cite{schmidtmobimm}.}

This technical note has at least one important theoretical implication.  {\em Benson et al.} \cite{Benson_arbitrary} suggested that the reactive-RWPT method, when combined with this mass transfer method, could partition the diffusion/dispersion process in any way that the physics demand. \aside{Imagine two systems with total molecular diffusion plus hydrodynamic dispersion of $D_{mol}+D_H=10^{-3}$.  One system with $D_{mol}=10^{-6}$ would have more mass transfer between nearby particles, and greater overall reaction rates, than a system with $D_{mol}=10^{-9}$, even though the hydrodynamic dispersion $D_H\approx10^{-3}$ would spread the species in nearly exactly the same way.  In fact, as $D_{mol}\rightarrow 0$, a simulation would revert to unreactive, conservative components.} So, if the dispersion tensor is thought of as a combination of velocity contrasts that promote spreading but not mixing, on top of smaller-scale mixing processes, then the reactive-RWPT method can very simply and separately perform true mixing (by mass transfer shown here) and macro-scale spreading via random walks.  In this way the reactive-RWPT method is solving a different equation than any Eulerian method.  Those methods cannot distinguish between the various components of the dispersion tensor $\vec D$.  To be more specific, the dispersion tensor is often assumed to follow \cite{Bear} ${\vec D} = (D_{\text{mol}}+\alpha_T\Vert \vec v \Vert){\vec I} +(\alpha_L  -\alpha_ T)\frac {\vec v \vec v^T}{\Vert \vec v \Vert}$,  where $D_{\text{mol}}$ is molecular diffusion, $\alpha_T < \alpha_L$ are transverse and longitudinal dispersivity, and $\vec v$ is a velocity column vector. {\em Cirpka and Werth, et al.} \cite{Cirpka_1999,Werth_focus} reinforce the view of {\em Gelhar et al.} \cite{Gelhar1979,Gelhar_1983}, who suggested that the first term (isotropic molecular diffusion plus smaller-scale transverse dispersion) truly represents a mixing process, while the addition of longitudinal dispersion accounts for velocity variations (hence a spreading process). Our method can separately simulate the smaller-scale mixing between particles (the first term) by the mass transfer algorithms shown here.  Particle separation, as by sub-grid velocity variations, can be separately handled by random walks.

\section*{References}

\bibliography{reaction_proposal}

\end{document}